\documentclass[a4paper,fleqn]{cas-dc}
\makeatletter
\expandafter\def\csname __first_footerline:\endcsname{%
  {\small\sffamily
  \ifnum\theblind>0\relax
  \else
    \csname __short_authors:\endcsname
  \fi}%
}
\makeatother
\usepackage[numbers]{natbib}
\usepackage{multirow}
\newcommand{\proposedmethod}{PHKT\xspace}
\def\tsc#1{\csdef{#1}{\textsc{\lowercase{#1}}\xspace}}
\tsc{WGM}
\tsc{QE}
\tsc{EP}
\tsc{PMS}
\tsc{BEC}
\tsc{DE}

\begin{document}
\let\WriteBookmarks\relax
\def\floatpagepagefraction{1}
\def\textpagefraction{.001}

\shorttitle{PHKT:Personalized Dynamic Hypergraph-enhanced KAN-Transformer for Multi-behavior Sequential Recommendation}
\shortauthors{R. Du et~al.}

\title[mode = title]{PHKT:Personalized Dynamic Hypergraph-enhanced KAN-Transformer for Multi-behavior Sequential Recommendation}

\author[1]{Ruijie Du}
\ead{duruijie@hdu.edu.cn}

\author[2]{Hao Chen}
\ead{haoc@hdu.edu.cn}

\author[2]{Xin Zhang}
\cormark[1]
\ead{zhangxin@hdu.edu.cn}

\author[2,3]{Dongjing Wang}
\cormark[2]
\ead{dongjing.wang@hdu.edu.cn}

\author[4]{Ze Zhang}
\ead{zhangzebut@163.com}

\author[5]{Xudong Shen}
\ead{hzshenxudong@corp.netease.com}

\author[5]{Runze Wu}
\ead{wurunze1@corp.netease.com}

\author[2]{Dongjin Yu}
\ead{yudj@hdu.edu.cn}

\affiliation[1]{
    organization={Hangzhou Dianzi University ITMO Joint Institute, Hangzhou Dianzi University},
    city={Hangzhou},
    postcode={310018},
    country={China}
}

\affiliation[2]{
    organization={School of Computer Science and Technology, Hangzhou Dianzi University},
    city={Hangzhou},
    postcode={310018},
    country={China}
}

\affiliation[3]{
    organization={Sanya Traditional Chinese Medicine Hospital},
    city={Sanya},
    postcode={572000},
    country={China}
}

\affiliation[4]{
    organization={Combat Support College, Rocket Force University of Engineering},
    city={Xi'an},
    postcode={710038},
    country={China}
}

\affiliation[5]{
    organization={Fuxi AI Lab, NetEase Games, NetEase Inc.},
    city={Hangzhou},
    postcode={310052},
    country={China}
}

\cortext[cor1]{Corresponding author. Tel.: +86 17681839286.}
\cortext[cor2]{Corresponding author. Tel.: +86 13732205613.}

\begin{abstract}
In multi-behavior recommendation, auxiliary behaviors such as clicks, add-to-cart, and purchases can provide richer supervisory information for predicting target behaviors. Although existing graph/hypergraph methods are capable of modeling high-order relationships among users, items, and behaviors, they still have limitations in heterogeneous semantics, user-specific weighting, and sequence dependency modeling. While standard Transformers excel at sequence modeling, their shared feedforward mapping struggles to accommodate the differentiated requirements of heterogeneous latent patterns in multi-behavior scenarios. To address this, this paper proposes the Personalized Hypergraph-enhanced Kolmogorov-Arnold Network Transformer (PHKT). Specifically, we design a personalized dynamic hypergraph module that performs behavior-aware weighting of item similarities based on users’ historical behavior sequences to capture user-specific heterogeneous high-order relationships; meanwhile, a Transformer is used as the temporal backbone to model the evolution of short- and long-term preferences, and the KAN is introduced to replace the traditional MLP in the feedforward network to enhance fine-grained modeling capability for nonlinear responses to different latent patterns. Experiments on three real datasets—Tmall, RetailRocket, and IJCAI show that PHKT consistently outperforms 9 strong baseline models across multiple evaluation metrics, demonstrating its effectiveness in multi-behavior preference modeling and target behavior prediction.
\end{abstract}



\begin{keywords}
Sequential Recommendation \sep Multi-Behavior Modeling \sep Hypergraph \sep Kolmogorov-Arnold Network \sep Interpretability
\end{keywords}

\maketitle

\section{Introduction}
With the explosive growth of online information and services (information overload), recommender systems~\cite{li2024recent,zhang2025multivariate,wang2025graph} have become indispensable components of digital platforms, helping users overcome information overload problems navigate through massive item collections and make informed decisions. These systems play a pivotal role across diverse domains, such as e-commerce, entertainment~\cite{yu2024mhaner}, and social media~\cite{yang2024cascading,wang2023multi}, generating substantial business value while enhancing user experience.
Traditional recommender systems primarily rely on static user-item interactions, treating each user action as an independent event. However, this approach overlooks a crucial aspect of user behavior: the temporal dynamics and sequential patterns that characterize real-world interactions. Users' preferences naturally evolve over time, and their current interests are often strongly influenced by their recent interaction history. For instance, a user's purchase decision may follow a distinct behavioral chain: browsing similar items, adding products to cart, and comparing alternatives before making the final selection.

\begin{figure}
  \centering
  \includegraphics[
    width=0.45\textwidth,
    height=0.45\textheight,
    keepaspectratio
  ]{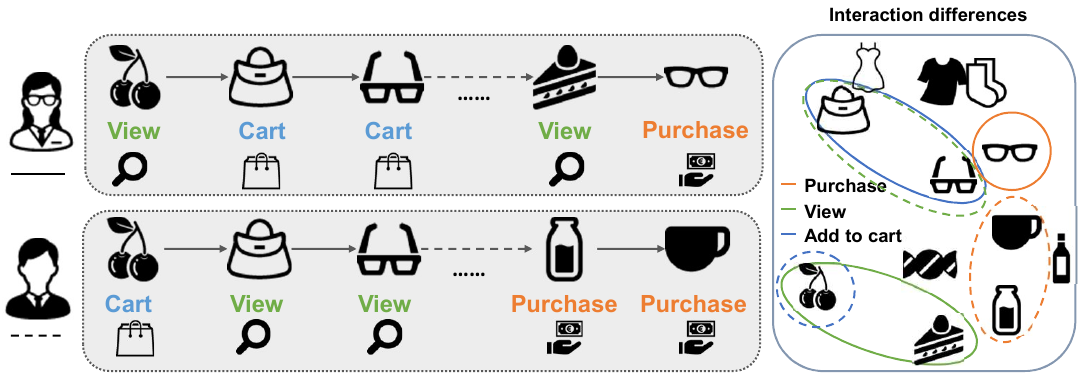}
  \caption{Distinct users’ interaction trajectories and behavioral differences in action patterns: a visualization of multi-behavior dynamic sequences for sequential recommendation.}
  \label{figure2}
\end{figure}

In order to better capture this behavioral dynamics, multi-behavior sequence recommendation has become an important research paradigm in modern online service platforms. They utilize users' multi-dimensional behavior sequences (including browsing, adding to cart, purchasing, and favoriting) to predict potentially interesting products. Although methods based on Transformer ~\cite{vaswani2017attention} and graph neural networks ~\cite{feng2019hypergraph,pan2025adaptive,liao2024hypergraph}. have demonstrated strong effectiveness in this task, they still face several core limitations: static hypergraphs are difficult to capture personalized heterogeneous dynamic dependencies, and the unified feedforward mapping of standard Transformers is hard to finely adapt to the differential semantic responses of patterns in multi-behavior heterogeneous scenarios.

\begin{figure}
  \centering
  \includegraphics[
    width=0.45\textwidth,
    height=0.45\textheight,
    keepaspectratio
  ]{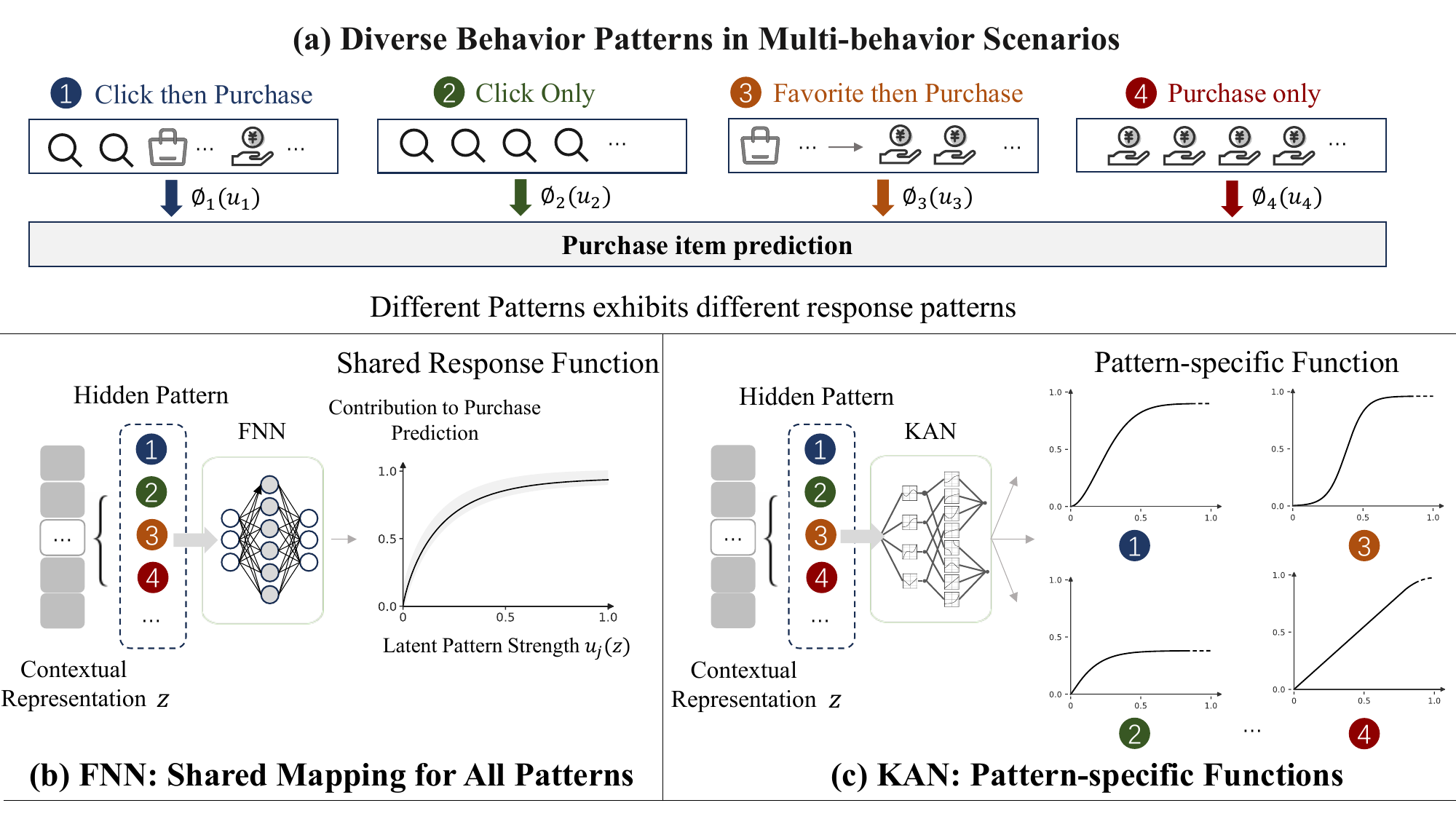}
  \caption{(a) Context representations based on self-attention encode the semantic information of multiple latent behavior patterns $\phi_j(\cdot)$. Different latent patterns correspond to different contribution response functions. As the pattern intensity changes, each latent pattern produces different contribution responses. (b) A standard feedforward neural network shares the same fixed activation template across all latent behavior patterns, so it can only perform a uniform, compromised fit. (c) KAN allows different latent behavior patterns to learn different response functions, achieving more fine-grained pattern-specific modeling.}
  \label{figure3}
\end{figure}

\textbf{Limitations of Static Hypergraph Modeling}. 
As illustrated in Fig.\ref{figure2}, user interaction sequences exhibit complex behavioral heterogeneity, where different interactions encode multi-scale preference semantics, such as short-term exploratory behaviors (impulses) and long-term preference patterns (inclinations). Existing methods often model such multi-behavior contexts under homogeneous assumptions, making it difficult to characterize personalized heterogeneous dependencies and their dynamic evolution. Although graph (hypergraph)-based frameworks offer a natural way to express high-order relationships, they usually assign static weights to (hyper-)edges, and cannot adapt to user-specific behavioral semantics. Therefore, they fundamentally fail to disentangle these multi-scale dependencies, particularly facing challenges in capturing: (1) dynamic transitions between heterogeneous behaviors (e.g., ``view–collect–purchase'' transition chains), (2) user-specific personalized interaction patterns, and (3) the high-order complex dependencies beyond simple bipartite structures.

\textbf{Limitations of Unified Feedforward Mapping}.
In multi-behavior recommendation, historical interactions processed by self-attention form a sequence of hidden representations that encodes rich contextual dependencies and high-order semantic relations. We refer to the high-order semantic units jointly induced by behavior order, behavior frequency, and contextual relations as \emph{latent behavioral patterns}, and define their prominence within the sequence as \emph{behavior intensity}, which is mainly reflected by the occurrence frequency of a pattern in interactions related to similar items.

Taking the latent behavioral patterns illustrated in Fig.~\ref{figure3}(a) as examples, we can further understand the different roles that different patterns play in target purchase prediction. For instance, when similar items in the history are associated with repeated clicks together with repeated purchases, the corresponding pattern usually exerts a strong promoting effect on purchase prediction. In contrast, when similar items in the history are only associated with repeated clicks without further conversion to purchase, the contribution of this pattern to the prediction may still increase with pattern strength, but its growth is usually slower and tends to saturate more easily in the high-intensity region.

These observations indicate that the contributions of different latent behavioral patterns to the prediction of the target purchase sample do not follow a unified response form. Instead, they are better characterized by pattern-specific nonlinear response functions, as illustrated in Fig.~\ref{figure3}(c).

However, in the classical Transformer, the standard feedforward layer adopts a unified shared mapping
\begin{equation}
f_{\mathrm{FFN}}(\mathbf z)=\mathbf W_2\,\sigma(\mathbf W_1\mathbf z+\mathbf b_1)+\mathbf b_2.
\end{equation}

This means that although the hidden representations after self-attention already encode multiple latent behavioral patterns, the standard FFN still forces these patterns to share the same fixed activation template in the nonlinear stage. Therefore, when the contribution--response relations of different latent patterns differ substantially, the standard FFN is more likely to learn a compromise mapping within a unified function family, rather than finely fitting the true contribution of each pattern to purchase prediction, as illustrated in Fig.~\ref{figure3}(b).

To address the aforementioned challenges, this paper proposes PHKT (Personalized Dynamic Hypergraph enhanced Kolmogorov–Arnold Network Transformer), a unified modeling framework for multi-behavior sequence recommendation. This model aims to collaboratively model the high-order structural relationships, temporal dependencies, and complex nonlinear preference patterns in multi-behavior interactions, thereby more comprehensively characterizing the evolution of user interests and the mechanisms of purchase conversion.

First, PHKT constructs a Personalized Hypergraph with Dynamic Weights (PHDW) module. This module extracts behavior-aware coefficients based on users’ historical interaction sequences and dynamically reweights the item similarity matrix accordingly to form a user-level personalized hypergraph structure. With this design, the model can simultaneously capture heterogeneous user features in multi-behavior interactions and high-order relationships among items, enhancing its capability to model complex behavioral dependency structures.

Subsequently, considering that the hypergraph branch alone is difficult to adequately represent the temporal dependencies and preference evolution in behavior sequences, this paper further introduces a Transformer as the backbone for temporal modeling to complement the shortcomings of structural modeling in characterizing sequence dynamics. At the same time, to mitigate the limitations caused by standard feedforward networks using a unified shared mapping, we replace the feedforward network in the Transformer with a KAN, using an adaptively weighted combination of radial basis functions instead of the traditional MLP transformation. As shown in Figure~\ref{figure3}(c), this design can better capture the contribution-response relationship of heterogeneous latent behavior patterns under different intensity modes, thereby enhancing the fine-grained modeling capability for multi-behavior semantic heterogeneity and complex purchase contribution relationships.

Extensive experiments on three real-world datasets (Tmall, RetailRocket, and IJCAI) demonstrate that \proposedmethod significantly outperforms existing baselines, including several state-of-the-art (SOTA) methods, particularly in modeling complex behavioral trajectories. Furthermore, ablation studies validate the effectiveness of our KAN-Transformer and personalized dynamic weight hypergraph fusion mechanisms.
In summary, our main contributions are as follows:
\begin{itemize}
     \item  We propose a unified multi-level architecture, Personalized Dynamic Hypergraph-enhanced Kolmogorov Arnold Network (KAN)-Transformer, for multi-behavior sequential recommendation. 
     \item We design a personalized hypergraph with a dynamic weight fusion mechanism for heterogeneous behavior modeling, which explicitly captures high-order, user-specific heterogeneous relationships across diverse user behavior patterns and enhances the model's ability to discern user preferences.
     \item We designed a KANT architecture to replace the MLP-style nonlinear transformation in the Transformer feed-forward network with a mapping based on adaptively weighted radial basis functions, thereby enhancing the model's fine-grained modeling ability of the heterogeneous responses between latent pattern semantics and target purchase prediction.
     \item We empirically demonstrate the effectiveness of \proposedmethod on three real-world datasets, providing novel architectural insights for sequential recommendation tasks.
\end{itemize}

The remainder of this paper is organized as follows. Section 2 reviews related work on sequential recommendation and related techniques. Section 3 gives the problem formalization and Section 4 details the proposed \proposedmethod architecture, including the design of the KAN-Transformer and the personalized hypergraph with dynamic weights. Section 5 describes the experimental setup, and presents and discusses the experimental results, including ablation studies and analysis of model components. Finally, Section 6 concludes the paper and outlines directions for future research.





\section{Related Work}
\subsection{Sequential Recommendation}
Early approaches for sequential recommendation approaches primarily relied on Markov chains to model user transitions among items. For example, Factorizing Personalized Markov Chains (FPMC)~\cite{rendle2010factorizing}integrates matrix factorization with Markovian processes, jointly capturing both sequential dependencies and individual user preferences. However, these methods are inherently constrained by the first-order Markovian assumption, which limits their ability to model long-range dependencies in user behavior sequences.

The introduction of recurrent neural network (RNN)-based frameworks marked a significant advancement. For instance, GRU4Rec~\cite{hidasi2015session} uses gated recurrent units (GRU) to process variable-length sequences, achieving better performance than Markov chain models in capturing complex temporal dynamics. Subsequently, Caser~\cite{tang2018personalized} applies convolutional neural networks to model both horizontal and vertical patterns within user interaction sequences, thereby refining the characterization of short- and long-term user preferences.

More recently, Transformer-based architectures have revolutionized sequential recommendation by leveraging the self-attention mechanism. SASRec~\cite{kang2018self} is a pioneering work that applies self-attention to capture long-range dependencies in user sequences, effectively mitigating the gradient vanishing problem inherent in RNNs. Building upon this, BERT4Rec~\cite{sun2019bert4rec} introduces bidirectional context modeling and frames the recommendation task as masked item prediction.

Despite these advances, existing methodologies remain constrained by certain limitations. Most methods treat user interactions as homogeneous events, lacking the ability to explicitly model heterogeneous relationships across different operation types—such as browsing, adding to cart, or purchasing~\cite{li2019multi}. Recent efforts, including LS-TGNN~\cite{ou2025ls} and DMI-GNN~\cite{lv2025dynamic}, have attempted to address these issues by incorporating auxiliary information or graph-based structures; however, effectively capturing the full spectrum of user interaction diversity remains challenging.

\subsection{Multi-behavior Recommendation}
Within the field of multi-behavior recommendation, existing studies can generally be grouped into three technical directions, namely graph-structured modeling, sequence-aware modeling, and generative modeling~\cite{liao2025fuzzy,yu2026multi}.

Among them, graph-based methods mainly focus on explicitly modeling the dependency structure among different behavior types. MB-GCN~\cite{jin2020multi} constructs a unified graph to represent multi-behavior interactions and models the varying influence of auxiliary behaviors on the target behavior, thereby improving the utilization of heterogeneous feedback. More recently, Li et al.~\cite{li2024hypergraph} further incorporate hypergraph modeling into multi-behavior sequential recommendation by constructing hypergraphs to capture high-order multi-behavior dependencies and fusing hypergraph information with sequential signals. However, although such methods are effective in modeling structural correlations, they still place primary emphasis on relational dependencies, while providing relatively limited characterization of the fine-grained semantic differences among heterogeneous behaviors.

Compared with graph-structured approaches, sequence-aware methods place greater emphasis on behavior order and the dynamic evolution of user preferences. GXNSRec~\cite{chen2026gxnsrec} introduces behavior-aware self-attention and graph cross networks to capture long-range dependencies and cross-behavior feature interactions in multi-behavior sequences. In addition, cd-MBRec~\cite{yan2025cd} explicitly models the commonality and diversity of multiple behaviors, which helps alleviate noise disturbance and ambiguous weight allocation in multi-behavior recommendation. Nevertheless, existing methods in this direction still face difficulties in jointly modeling sequential dynamics and heterogeneous behavioral semantics in a unified and fine-grained manner.

Generative methods attempt to characterize multi-behavior interactions from the perspective of behavior generation. MBGen~\cite{liu2024multi} formulates multi-behavior recommendation as a unified generative process by interleaving behavior and item tokens, and improves recommendation performance through autoregressive prediction of the next behavior and item. Nevertheless, although this generative paradigm is effective in modeling the dependency between behaviors and items, it still does not explicitly disentangle user-specific semantic relations among different behavior types, which may limit its fine-grained personalized modeling capability.

Overall, many existing methods still tend to treat user behavior sequences in an insufficiently fine-grained manner, or struggle to model the heterogeneous semantics of multiple behaviors together with their dynamic evolution. As a result, these methods often fail to fully capture the heterogeneity and dynamics inherent in user behaviors, which also highlights the urgent need for more sophisticated and powerful models that can jointly characterize heterogeneous behavioral patterns and their dynamic evolution.

\subsection{Fusion methods of KAN}
Kolmogorov--Arnold Networks (KANs)~\cite{DBLP:conf/iclr/LiuWVRHS0T25,somvanshi2025survey,liu2025weighted} have recently attracted increasing attention in recommendation systems as a promising nonlinear approximation paradigm. By leveraging learnable basis function compositions, KANs provide a flexible alternative to conventional neural architectures for modeling complex nonlinear relationships.

Recent studies have begun to explore the integration of KAN with recommendation models from different perspectives. For example, FourierKAN-GCF~\cite{xu2024fourierkan} incorporates Fourier series into graph-structured collaborative filtering, improving the adaptability of representation learning on high-dimensional graph data. In addition, CF-KAN~\cite{park2024cf} applies KAN to collaborative filtering under continual learning settings, focusing on alleviating catastrophic forgetting through parameter isolation and robust knowledge retention. These studies demonstrate the feasibility of combining KAN with recommendation frameworks and suggest its potential for enhancing nonlinear representation ability.

However, current applications of KAN in recommendation systems remain relatively limited and have not yet fully explored its potential. In particular, its advantages for modeling complex nonlinear relationships in multi-behavior recommendation scenarios have not been sufficiently investigated.

\subsection{Hypergraph Neural Network}
In complex data relationship modeling, hypergraphs have become an effective tool due to their ability to capture high-order correlations beyond traditional graph structures~\cite{feng2019hypergraph,pan2025adaptive}. For example, Feng et al.~\cite{feng2019hypergraph} proposed the Hypergraph Neural Network (HGNN) framework, which encodes high-order correlations in hypergraph structures through hyperedge convolution. As a generalized learning paradigm, HGNN enables the learning of latent representations that better reflect the high-order structural characteristics of real-world data.

In recommendation systems, hypergraph-based models~\cite{choi2025hypergraph,li2024hypergraph,zhou2026hypergraph} have also been introduced to better characterize complex user--item interaction patterns. As an early representative work, Xia et al.~\cite{xia2021self} modeled session-based data as a hypergraph and proposed a dual-channel hypergraph convolutional network to capture higher-order item transitions and enrich session representations. Building on this line of research, more recent studies have further extended hypergraph modeling to multi-behavior recommendation. For instance, Li et al.~\cite{li2024hypergraph} incorporated hypergraph-enhanced multi-interest learning into multi-behavior sequential recommendation, while Zhou et al.~\cite{zhou2026hypergraph} further enhanced recommendation by exploiting diverse higher-order interactions through a hypergraph mixture-of-experts framework.

However, most existing hypergraph-based recommendation methods are still limited by static hyperedge designs. In real-world scenarios, user behaviors and user--item relationships often exhibit dynamic evolutionary patterns, while existing static hypergraph structures are difficult to adapt to such behavior-driven dynamic associations, which may result in suboptimal performance. In addition, many existing methods do not explicitly model the high-order relationships across different behavior types (e.g., browsing, adding to cart, and purchasing). Since different behaviors often carry distinct semantic meanings and interconnections, neglecting such heterogeneity may lead to an insufficient understanding of user preferences and further reduce recommendation accuracy~\cite{yang2022multi}.


\section{Problem Formalization}
In this paper, we denote the set of items as $v_{i} \in \boldsymbol{\mathcal{V}}$, the set of behavior types as $t_{j} \in \boldsymbol{\mathcal{T}}$, and the set of users as $u \in \boldsymbol{\mathcal{U}}$.
The multi-behavior interaction sequence of a user $u$ is represented as $\boldsymbol{S}{u} = \left[\left(v{1}, t_{1}\right), \left(v_{2}, t_{2}\right), \ldots, \left(v_{L}, t_{L}\right)\right]$, where $L$ denotes the sequence length and each tuple $(v_{l}, t_{l})$ corresponds to the interaction of user $u$ with item $v_{l}$ through behavior type $t_{l}$ at the $l$-th position in the sequence.

Given a historical sequence $\boldsymbol{S}{u}$ of length $L$ for user $u$, the objective of multi-behavior sequential recommendation is to predict the next item-behavior that the user is most likely to interact with, considering both the sequential dependencies and the heterogeneity of different behavior types.

To represent the heterogeneous and sequential nature of user interactions, we adopt the joint embedding scheme from MBHT~\cite{yang2022multi}, which consists of three components:

(i) \textbf{Item embeddings} $E_{v} \in \mathbb{R}^{|\mathcal{V}| \times d}$, encoding the latent representation of each item;

(ii) \textbf{Behavior-type embeddings} $E_{t} \in \mathbb{R}^{|\mathcal{T}| \times d}$, capturing the semantics of different behavior types (e.g., click, add-to-cart, purchase);

(iii) \textbf{Position embeddings} $E_{p} \in \mathbb{R}^{L \times d}$, encoding the order information within the interaction sequence.

The final input embedding for the sequence is obtained by fusing these three components as follows:
\begin{equation}
\boldsymbol{X} = E_{v} \oplus E_{t} \oplus E_{p}
\end{equation}
where $\oplus$ denotes the fusion operation, typically implemented with element-wise addition or concatenation.

For clarity, we summarize the key notations used throughout this paper in Tab.\ref{table0}, which readers may refer to for quick reference.

\begin{table}[width=.9\linewidth,cols=4,pos=h]
\centering
\caption{Key Notations in This Paper}
\begin{tabular}{c|p{6cm}}
\hline
Symbol        & Description \\ \hline
$\boldsymbol{\mathcal{V}}$, $\boldsymbol{\mathcal{T}}$, $\boldsymbol{\mathcal{U}}$ & Sets of items, behavior types and users\\
$\boldsymbol{S}_{u}$ & Interaction sequence of user $u$\\
$E_{v}$, $E_{t}$, $E_{p}$ & Embeddings of items, behavior types and positions\\
$\boldsymbol{X}_{KAN-Trans}^{out}$ & Output of Transformer with KAN layer\\
$\boldsymbol{X}_{v}$ & Representation of item embedding\\
$\boldsymbol{\operatorname{Sim}}$ & Similarity matrix between items\\
$\mathbf{W}$ & Behavior based dynamic weight matrix calculated based on user historical interaction sequences\\
$\boldsymbol{G}$ & Hypergraph Propagation Matrix\\
$\boldsymbol{D}_{v}$ & Node Degree Matrix\\
$\boldsymbol{D}_{e}$ & Hyperedge Degree Matrix\\
$\boldsymbol{H}$ & Dynamically weighted hypergraph adjacency matrix\\
$\boldsymbol{X}_{G}^{out}$ & Output of Hypergraph Neural Network\\
$\boldsymbol{X}_{fused}$& Fusion of Hypergraph Neural Network and KAN Transformer Output\\ \hline
\end{tabular}
\label{table0}
\end{table}


\section{The Proposed Method-\proposedmethod}
Our work bridges the gap between hypergraph learning and behavior aware sequence modeling by introducing a new framework that integrates personalized dynamic weight hypergraph and KAN Transformer collaboration, which is a new attempt in the research of recommendation systems.

\begin{figure*}[t]
\centering
\includegraphics[width=1\textwidth]{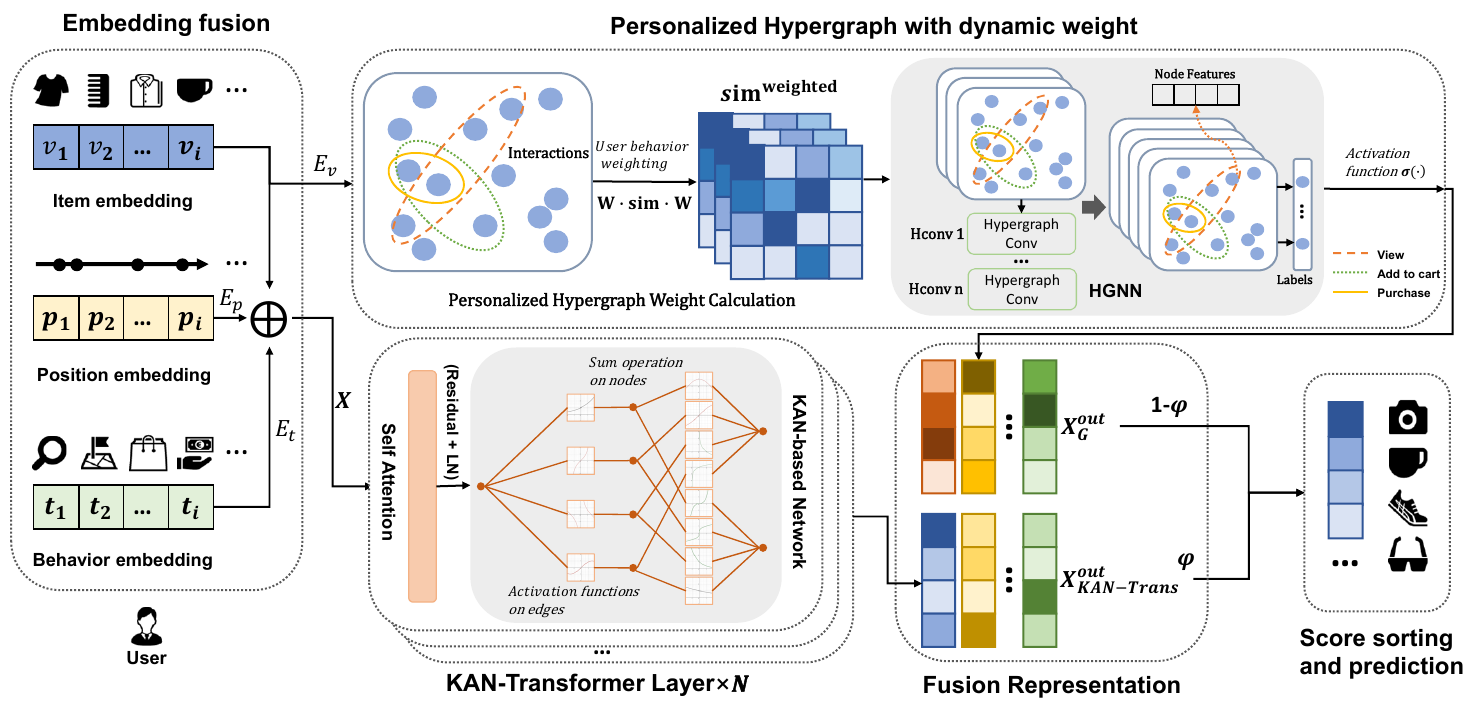} 
\caption{\proposedmethod first generates three types of embeddings based on the user's historical interaction information: item, location, and interaction behavior. Then, combining these three embeddings, it generates a fused embeddings $\boldsymbol{X}$ and inputs it into a Transformer embedded with KAN for feature extraction. At the same time, calculate the similarity matrix $\boldsymbol{\operatorname {Sim}}$ based on the item embedding, and use the user's historical behavior information to weight this similarity matrix to obtain $\boldsymbol{\operatorname {Sim}}^{\text {weighted }}$. Send the weighted hypergraph into HGNN to obtain personalized hypergraph output, and finally use its combined output with KANT learning to predict and sort the next transaction item.}
\label{figure1}
\end{figure*}

As is shown in Fig.\ref{figure1}, which presents the overall framework of the algorithm proposed in this paper. This section provides a formal definition of symbols and a statement on the multi-behavior recommendation problem in this chapter, laying a solid theoretical foundation for subsequent research. Subsequently, the key components were elaborated in detail. Firstly, it introduces how KANT layer serves as the perception core of feedforward network and how transformer encoders capture sequential dependencies. Then, the approaches of Personalized Hypergraphs with Dynamic Weight (PHDW)  was introduced, including hypergraph definition, dynamic hypergraph weight calculation, and the hypergraph convolution operation. Finally, the prediction layer and the loss function were elucidated while the process of generating project scores and optimizing model parameters to achieve accurate recommendations was detailed.

\subsection{Behavior-aware KAN-Transformer for representation learning}

As mentioned earlier, Traditional Transformers rely on feedforward networks for feature transformation, but their shared MLP structure struggles to finely capture heterogeneous response relationships in multi-behavior scenarios.. To address this, we introduce the Kolmogorov-Arnold Network (KAN) into the Transformer's feedforward network to reconstruct its main nonlinear mapping process. Specifically, we use KAN to construct independent radial basis functions for each input dimension and combine them through learnable weights to replace the original MLP-style transformation, thereby enhancing the model's ability to adapt to the semantic heterogeneity of potential patterns and their differentiated nonlinear response requirements in multi-behavior scenarios.
\subsubsection{Design concept of KANT}
The FNN(Feedforward Neural Network) structure of the original Transformer is defined as follows:
\begin{equation}
\operatorname{FNN}(x)=W_{2} \cdot \sigma\left(W_{1} \cdot x+\alpha_{1}\right)+\alpha_{2}
\end{equation}
which $x$ represents the input vector of the FNN layer, while $W$ and $\alpha$ are learnable parameters. And $\sigma$ is defined as the activation function within FNN. We replace one linear layer and the original activation in FNN with a KAN layer (KANT), whose basis functions are defined as follows:
\begin{equation}
B_{i}\left(x_{i}\right)=\left[b_{i, 1}\left(x_{i}\right), b_{i, 2}\left(x_{i}\right), \cdots, b_{i, k}\left(x_{i}\right)\right] \in \mathbb{R}^{K}
\end{equation}
Among these, \(b_{i,k}(\cdot)\) denotes the \(k\)-th radial basis function on the \(i\)-th input dimension, and \(K\) denotes the number of basis functions associated with that dimension. Considering that the response relationships of different latent patterns in multi-behavior scenarios usually exhibit pronounced local nonlinear characteristics, radial basis functions are more suitable for this setting due to their advantages in local modeling and smooth approximation. Subsequently, the outputs of the basis functions on each dimension are independently linearly projected:
\begin{equation}
z_{i}=B_{i}\left(x_{i}\right)\cdot \omega_{i},\quad \omega_{i} \in \mathbb{R}^{K\times d_{\text {out}}}
\end{equation}
The $\omega_{i}$ in the above formula represents the learnable parameters of basis function combinations in the KANT layer. Ultimately, by adding the outputs of all dimensions, the output vector is derived as follow:
\begin{equation}
K A N(x)=\sum_{i=1}^{d} z_{i}=\sum_{i=1}^{d} B_{i}\left(x_{i}\right) \cdot \omega_{i}
\end{equation}
In the specific implementation of KANT, we aggregate the responses of each basis function using learnable weights, without explicitly introducing the higher-order combination forms in the original KAN. This design preserves the adaptive nonlinear mapping characteristics of KAN while controlling additional parameter overhead and optimization difficulty, making it more suitable for recommendation scenarios. Based on this, the entire feedforward layer can be expressed as:
\begin{equation}
\boldsymbol{X}_{KAN-Trans}^{out}=\operatorname{Layernorm}(\operatorname{KAN}(x)+x)
\end{equation}
\subsubsection{Theoretical Analysis of KANT}
To further illustrate why this structure can provide more fine-grained modeling capability, let the contextualized representation generated by self-attention be denoted as
\begin{equation}
x = h_{\theta}(S_u) \in \mathbb{R}^{d}
\label{eq:contextualized_representation}
\end{equation}
where \(S_u\) denotes the user interaction sequence and \(h_{\theta}(\cdot)\) denotes the preceding encoding process. This representation is formed by the global context of the entire behavior sequence and therefore typically contains information related to different latent behavioral patterns.

In KANT, the \(q\)-th output channel can be written as
\begin{equation}
\operatorname{KAN}^{(q)}(x)=\sum_{i=1}^{d}\phi_{i,q}(x_i)
\label{eq:kant_output_channel}
\end{equation}
where \(\phi_{i,q}(\cdot)\) denotes the learnable response function from the \(i\)-th input dimension to the \(q\)-th output channel. Unlike the standard FFN, in which all dimensions share the same nonlinear template, the functions \(\phi_{i,q}(\cdot)\) in KANT are parameterized independently. Therefore, different dimensions are not constrained to follow the same response function form.

During the end-to-end training process, the prior encoder gradually organizes the semantic information of different latent behavioral patterns into different dimensions and their combined representations; meanwhile, KANT learns their respective independent response functions on each dimension through basis function combinations. As training progresses, these dimension-level functions collaboratively form an overall response mapping f(x) for different latent behavioral patterns. Due to differences in the activation structures corresponding to different patterns, the overall response function also exhibits pattern specificity, enabling the model to more finely characterize the heterogeneous behavioral pattern response relationships in multi-behavior recommendations.

In summary, the KANT layer preserves the nonlinear expansion mechanism of KAN by generating basis-function responses for each input dimension and combining them with learnable weights for fine-grained nonlinear modeling. Integrated into the Transformer feed-forward layer while retaining residual connections and LayerNorm, transforms the original MLP-style feed-forward network into a KAN-style mapping. This design enables the model to better fit heterogeneous nonlinear responses across latent patterns in multi-behavior scenarios, thereby capturing the fine-grained contributions of different pattern semantics to target purchasing behavior more effectively.

\subsection{Personalized Hypergraph with Dynamic Weight for Multi-behavior Modeling}
In the realm of hypergraph-based recommendation systems, most existing hypergraph based recommendation methods treat all user interactions as homogeneous entities within the hypergraph structure, which ignore a critical aspect that user behaviors are inherently diverse in terms of their semantic implications. Consequently, traditional methods fail to capture the subtle semantic differences from user preferences. In contrast, our method dynamically calculates weights based on the user's historical behavioral preferences, enhancing the influence of items under important interaction behaviors while reducing the impact of noise on user preference capture (e.g., viewing behaviors that may including accidental clicks, and the large quantity of viewing behavior can cause a lot of noise to interfere with the accuracy of the model), forming a hypergraph with enhanced dynamic behavior patterns.

To compensate for the lack of flexibility in modeling high-order behavioral relationships between different users $u$, we constructed a dynamic weighted hypergraph $G$. Formally, $G$ is characterized by an adjacency matrix $\boldsymbol{H} \in \mathbb{R}^{L \times E}$, and $V$ is the set of item nodes with length $L$. Each interactive behavior $t$ is assigned an initial weight $\alpha$, which is determined by the type of behavior: $\operatorname{bweight}(t)=\alpha$. In general, the weight of purchase behavior is the highest, and the weight of viewing behavior is lower than the weight of adding to cart. The initial weight vector for behavior types [view, add-to-cart, purchase, others] in our study is [0.3, 0.7, 1, 0.4]. This design simulates the differences in user preferences exhibited by each type of behavior, which is a prerequisite to address the limitations of traditional static hyperedge construction in the next step.

Then, we calculate the similarity between all items on the sequence embedding representation $\boldsymbol{X}_{v} \in \mathbb{R}^{L \times d}$, where $d$ denotes the dimension of the item embedding. This step aims to capture the potential relationship patterns implicitly reflected between user interaction items. Specifically, we adopt cosine similarity as the metric, defined as:
\begin{equation}
\boldsymbol{\operatorname{Sim}}=\operatorname{cosine}(\boldsymbol{X}_{v}, \boldsymbol{X}_{v})=\operatorname{normalize}(\boldsymbol{X}_{v}) \cdot \operatorname{normalize}(\boldsymbol{X}_{v})^{\top}
\end{equation}
Here, $\boldsymbol{\operatorname{Sim}} \in \mathbb{R}^{L \times L}$ represents the original similarity matrix, describing the latent semantic proximity between all pairs of elements. To further incorporate the behavioral context of each interaction, we use behavior-specific weights to reweight the similarity matrix:
\begin{equation}
\operatorname{\boldsymbol{\operatorname {Sim}}}^{\text {weighted }}=\mathbf{W} \cdot \boldsymbol{\operatorname{Sim}} \cdot \mathbf{W}
\end{equation}
\begin{equation}
\mathbf{W}=\operatorname{diag}\left(\operatorname{bweight}\left(t_{1}\right), \cdots\right., \operatorname{bweight} \left.\left(t_{L}\right)\right)
\end{equation}
This operation allows the similarity between each pair of items to be jointly influenced by the behavioral semantics at both ends, thereby enhancing similarities associated with high-value behaviors while suppressing the interference introduced by noisy behaviors. For each item $v$, we select $k$ of the most similar items in the weighted similarity matrix to form a hyperedge $\mathbf{e}_{i}$:
\begin{equation}
\boldsymbol{e}_{i}=\{\boldsymbol{v}\} \cup \operatorname {TopK}\left(\mathbf{Sim}_{\boldsymbol{v},:}^{\text {weighted}}, k\right)
\end{equation}
Then the hyperedge set can be represented as:
\begin{equation}
\boldsymbol{\varepsilon}^{u}=\left\{\mathbf{e}_{1}, \mathbf{e}_{2}, \cdots, \mathbf{e}_{L}\right\}
\end{equation}
Each column in the hypergraph adjacency matrix $\boldsymbol{H}$ corresponds to an $\mathbf{e}_{i}$, and the row indicates whether each item belongs to the hyperedge and is weighted according to the value of $\mathbf{Sim}^{\text {weighted }(u)}$.
With HGNN (Hypergraph Neural Network) propagation mechanism, we use the previously established hypergraph for feature learning. $\boldsymbol{D}_{\boldsymbol{v}}=\operatorname{diag}(\boldsymbol{H} \cdot \mathbf{1})$ represents the node degree matrix, $\boldsymbol{D}_{\boldsymbol{e}}=\operatorname{diag}\left(\mathbf{1}^{\top} \cdot \boldsymbol{H}\right)$ is the hyperedge degree matrix, and the propagation of the hypergraph is:
\begin{equation}
\boldsymbol{G}=\boldsymbol{D}_{v}^{-1} \boldsymbol{H} \boldsymbol{D}_{e}^{-1} \boldsymbol{H}^{\top}
\end{equation}
\begin{equation}
\boldsymbol{X}_{G}^{out}=\sigma\left(\boldsymbol{G} \cdot \boldsymbol{X}_{in} \cdot \boldsymbol{W}_{\boldsymbol{h}}\right)
\end{equation}
where $\boldsymbol{W}_{h}$ is the linear transformation matrix that can be learned by HGNN, and $\sigma$ is the activation function in HGNN.

By utilizing hypergraph convolution under behavior specific dynamic weights, our method captures the dynamic behavior patterns of users and advances the personalized expression of the model.
This dual innovation enables our framework to simultaneously address personalized behavioral heterogeneity (through the fusion of dynamically weighted hypergraphs and KANT), thus pioneering new approaches in multi behavior recommendation.

\subsection{Prediction and Recommendation}

Our method adopts the Masked Item Prediction strategy when predicting, that is, randomly select a part of the items to be replaced by MASK, and finally achieve the purpose of predicting the real items in these positions. This method has been widely used in sequence recommendation.

Combining KAN-Transformer and hypergraph representation, we get:
\begin{equation}
\boldsymbol{X}_{fused}=\varphi \cdot \boldsymbol{X}_{KAN-Trans}^{out}+(1-\varphi) \cdot \boldsymbol{X}_{G}^{out}
\end{equation}
The fusion coefficient $\varphi$ is a scalar learned by the attention mechanism. For each masked position $(s, l) \in M$, take out its representation vector $\boldsymbol{x}_{s, l}$, and then do a dot product with all item embeddings to get the predicted logits:
\begin{equation}
\boldsymbol{o}_{s, l}=\boldsymbol{x}_{s, l} \cdot \boldsymbol{E}^{\top} \in \mathbb{R}^{N}
\end{equation}
Among them, $\boldsymbol{E}^{\top} \in \mathbb{R}^{\boldsymbol{N} \times \boldsymbol{d}}$ is the embedding matrix of all items, and $N$ is the number of items. That is, the output of the encoder is mapped to the item set through the fully connected layer to obtain the predicted item score.

The standard cross-entropy loss is used here, which is calculated only on the mask position. $y_{s, l} \in\{1, \cdots, N\}$ is the real item ID in the mask position $(s, l)$, and the loss function is:
\begin{equation}
\operatorname{Loss}_{C E}=-\frac{1}{|M|} \sum_{(s, l) \in M} \log \left(\frac{\exp \left(o_{s, l, y_{s, l}}\right)}{\sum_{j=1}^{N} \exp \left(o_{s, l, j}\right)}\right)
\end{equation}
where $\boldsymbol{o}_{s, l, j}$ is the score of item $j$ and $M$ is the total number of masks.



\section{Experiments}
\subsection{Evaluation Setting}
In this subsection, we introduce the evaluation settings in details, including the datasets, implementation details, and baselines.
\subsubsection{Dataset}
As shown in Tab.\ref{table1} , we evaluated our method \proposedmethod on three real-world multi-behavior datasets:
\begin{itemize}
    \item \textbf{Retail Rocket}:This dataset, curated by Retail Rocket over a four-month period, documents three categories of user behaviors: the primary shopping action, complemented by supportive behaviors such as page browsing and cart additions.
    \item \textbf{IJCAI}: Released in the IJCAI competition for buyer behavior re-forecasting, this dataset shares the same interaction types as the Tmall dataset. However, variations exist across datasets in terms of average sequence length and user-item interaction density, offering distinct evaluation benchmarks.
    \item \textbf{Tmall}: The dataset is derived from Tmall, a leading e-commerce platform in China, and encompasses four distinct types of user-item interactions: the core target behavior (purchase), alongside auxiliary behaviors, including adding to favorites, cart additions, and page views.
\end{itemize}

For all three datasets, a unified preprocessing strategy was adopted before model training. Specifically, users were automatically split into training and test sets at a ratio of 8:2, and when necessary, the training set was further divided into training and validation subsets using the same ratio. Moreover, to alleviate data sparsity and improve data quality, users with fewer than 5 interaction records and items with fewer than 10 interactions were removed.

\begin{table}[htbp]
\centering
\caption{Dataset Statistics}
\label{table1}
\resizebox{\linewidth}{!}{
\begin{tabular}{p{0.2\linewidth}|p{0.08\linewidth}p{0.12\linewidth}p{0.18\linewidth}p{0.23\linewidth}}
\hline
Dataset & \multicolumn{1}{c}{\#(users)} & \#(items) & \#(behaviors) & \#(interactions) \\
\hline
\textbf{RetailRocket} & 235,061 & 1,407,580 & 3 & 2,756,101 \\
\textbf{IJCAI} & 1,090,390 & 424,170 & 4 & 54,925,330 \\
\textbf{Tmall} & 4,162,024 & 987,994 & 4 & 100,150,807 \\
\hline
\end{tabular}
}
\end{table}

\subsubsection{Implementation Details}
All experiments in this study were conducted on a Windows server equipped with an NVIDIA GeForce RTX 5060 Ti graphics card. The experimental environment is based on the Python platform, and the PyTorch framework (Python version 3.8.0) is used to implement the proposed method and compare it with the baseline.

We adopt standard metrics for sequential recommendation, including HR@5/10, NDCG@5/10, and MRR@5. HR measures whether the target item appears in the top-$k$ results, NDCG evaluates ranking quality by considering both relevance and position, and MRR measures the reciprocal rank of the first relevant item.

\subsubsection{Baselines}
To assess the efficacy of our proposed approach, we conduct comparisons against a set of state-of-the-art algorithms, which are categorized as follows:

\textbf{General sequence recommendation}
\begin{itemize}
\item \textbf{GRU4RecCPR}~\cite{chang2024copy,tan2016improved} This method augments the GRU4Rec sequential recommendation architecture with Softmax-CPR, enabling the model to better handle repeated-item recommendation by improving the output softmax layer and balancing the tendency to copy items from the interaction history against the prediction of novel items.

\item \textbf{SASRecCPR}~\cite{chang2024copy,kang2018self} This method extends SASRec with Softmax-CPR, allowing the self-attention-based sequential recommender to better distinguish when to copy previously interacted items and when to recommend new ones, thereby improving repeated-item recommendation.

\item \textbf{FEARec}~\cite{du2023fearec} This method introduces a frequency-enhanced hybrid attention mechanism that jointly models user behavior sequences in both the time and frequency domains, so as to capture long-range dependencies, periodic patterns, and high-frequency information more effectively in sequential recommendation.
\end{itemize}

\textbf{Graph-based Sequential Recommendation Systems}
\begin{itemize}
\item \textbf{SURGE}~\cite{chang2021surge} This method reconstructs user behavior sequences into item--item interest graphs via metric learning and applies graph neural networks to model high-order item transitions, thereby capturing users' dynamic core interests from noisy sequential behaviors. 

\item \textbf{MAERec}~\cite{ye2023maerec} This method builds a global item transition graph from user sequences and introduces a graph masked autoencoder to adaptively reconstruct informative graph structures, so as to enhance sequential recommendation under data sparsity and noise. 

\item \textbf{SelfGNN}~\cite{liu2024selfgnn} This method constructs short-term interaction graphs based on time intervals and employs graph neural networks together with self-supervised learning to capture cross-user collaborative signals, while improving robustness against noisy short-term 
\end{itemize}

\textbf{Multi-Behavior Recommendation Models}
\begin{itemize}
\item \textbf{MBHT}~\cite{yang2022multi} This method integrates a multi-scale Transformer with a hypergraph neural network to jointly capture short-term and long-term cross-type behavior dependencies, where low-rank self-attention is used to encode behavior-aware sequential patterns at different temporal granularities and the hypergraph module is introduced to model global high-order dependencies among multiple behaviors.
\item \textbf{PBAT}~\cite{su2023personalized} This method proposes a personalized behavior-aware Transformer for multi-behavior sequential recommendation, introducing a personalized behavior pattern generator to learn dynamic and discriminative behavior patterns and a behavior-aware collaboration extractor to incorporate both behavioral and temporal impacts into sequential collaborative transitions.
\item \textbf{MBSRec}~\cite{elsayed2024multi} This method presents a lightweight multi-behavior sequential recommendation framework that employs multi-head self-attention to capture the dependencies among heterogeneous historical interactions, while further using a weighted binary cross-entropy loss to achieve finer behavior-aware prediction.
\end{itemize}

\subsection{Comparison and Analysis}
The experimental results of the proposed method \proposedmethod, as reported in Tables~\ref{table2}--\ref{table4}, demonstrate its effectiveness across all three real-world datasets, i.e., RetailRocket, Tmall, and IJCAI.

Across all datasets, the proposed method \proposedmethod approach consistently achieves the best overall results in terms of HR@5, HR@10, NDCG@5, NDCG@10, and MRR@5, indicating the consistent effectiveness and robustness of our model in diverse real-world recommendation scenarios.

On the \textbf{IJCAI} dataset, \proposedmethod reaches an HR@10 of 0.583, an NDCG@10 of 0.495 and an MRR of 0.460. Compared with the second-best baseline MBHT, our model obtains performance gains of 49.9\%, 56.7\% and 42.9\% respectively. Such substantial performance improvements validate the superior recommendation capability of \proposedmethod on sparse and complex interactive scenarios of the IJCAI dataset.

For the \textbf{Tmall} dataset, the \proposedmethod model yields HR@10, NDCG@10 and MRR values of 0.440, 0.359 and 0.326. It surpasses the competitive baseline MBSRec by 26.1\%, 26.0\% and 50.2\% in corresponding metrics. Although the overall improvement is relatively moderate compared with that on the IJCAI dataset, \proposedmethod still maintains a clear performance advantage over other competitors.

In terms of the \textbf{RetailRocket} dataset, \proposedmethod achieves outstanding results with HR@10 of 0.952, NDCG@10 of 0.911 and MRR of 0.899. Relative to the suboptimal model MBHT, the performance increments are 4.2\%, 1.2\% and 0.2\%. Even though the margin is limited, these steady enhancements are realized on the basis of the already saturated performance of existing methods, which proves that \proposedmethod can further refine recommendation results in high-performance scenarios.

By comparing the experimental results on three datasets, we observe that all models present the best overall performance on RetailRocket. This phenomenon can be attributed to the regular user behavioral patterns and explicit item correlation within this dataset. In contrast, relatively lower results are reported on IJCAI and Tmall. The underlying reason lies in severe data noise, category imbalance and drastic dynamic changes in user preference in these two realistic datasets.

Notably, \proposedmethod maintains the optimal results on all three datasets with distinct data characteristics. It strongly verifies the outstanding generalization and robustness of our method, enabling it to adapt to diverse data distributions and complex application environments in real-world recommendation systems.

From the perspective of technical evolution, experimental results on three public datasets consistently demonstrate that graph neural networks outperform conventional sequential models. 
On the IJCAI dataset, the proposed SelfGNN achieves an HR@10 of 0.217, an NDCG@10 of 0.205, and an MRR of 0.178. Compared with the classical sequential baseline SASRecCPR, which obtains 0.140 for HR@10, 0.126 for NDCG@10 and 0.106 for MRR, the above three metrics are improved by 55.0\%, 62.7\% and 67.9\%, respectively. Such significant performance gains fully validate the effectiveness of graph structure modeling for sequential recommendation.

For the Tmall dataset, SelfGNN reaches 0.190 in HR@10, 0.185 in NDCG@10 and 0.153 in MRR. By contrast, SASRecCPR only yields an HR@10 of 0.163, an NDCG@10 of 0.119 and an MRR of 0.103. Relative to SASRecCPR, SelfGNN obtains consistent increments of 16.6\% in HR@10, 55.5\% in NDCG@10 and 48.5\% in MRR, which further verifies the inherent superiority of graph-based modeling methods.
On the RetailRocket dataset, the HR@10, NDCG@10 and MRR scores of SelfGNN are 0.896, 0.857 and 0.833, while the corresponding results of SASRecCPR are 0.725, 0.659 and 0.623. SelfGNN surpasses the baseline by 23.6\%, 30.0\% and 33.7\% across all evaluation metrics, presenting substantial and stable performance improvements.

Furthermore, multi-behavior recommendation models (i.e., MBSRec and MBHT) achieve outstanding results on all three datasets. Especially on the IJCAI and Tmall datasets, the MRR values of MBSRec are 0.322 and 0.217, leading to remarkable increases of 80.9\% and 41.8\% compared with SelfGNN. These evident performance improvements indicate that auxiliary multi-behavior information enables the model to capture fine-grained user preference intensity and complex interaction patterns more accurately.

In summary, the comprehensive experimental results illustrate the effectiveness of the proposed approach \proposedmethod for multi-behavior sequential recommendation, showing its value for deployment in various real-world industrial contexts.

\begin{table}[htbp]
\centering
\caption{Experimental results of the proposed approach \proposedmethod and baselines on the dataset of RetailRocket}
\resizebox{\linewidth}{!}{
\begin{tabular}{l|lllll}
\hline
RetailRocket & HR@5 & HR@10 & NDCG@5 & NDCG@10 & MRR@5 \\
\hline
FEARec & 0.714 & 0.785 & 0.706 & 0.777 & 0.689 \\
SASRecCPR & 0.701 & 0.725 & 0.621 & 0.659 & 0.623 \\
GRU4RecCPR & 0.752 & 0.806 & 0.614 & 0.683 & 0.630 \\
\hline
MAERec & 0.868 & 0.887 & 0.842 & 0.864 & 0.841 \\
SelfGNN & 0.883 & 0.896 & 0.821 & 0.857 & 0.833 \\
SURGE & 0.865 & 0.898 & 0.862 & 0.879 & 0.859 \\
\hline
PBAT & 0.874 & 0.901 & 0.857 & 0.895 & 0.844 \\
MBSRec & 0.895 & 0.908 & 0.873 & 0.897 & 0.886 \\
MBHT & \underline{0.906} & \underline{0.914} & \underline{0.897} & \underline{0.900} & \underline{0.897} \\
\hline
\textbf{PHKT} & \textbf{0.926} & \textbf{0.952} & \textbf{0.902} & \textbf{0.911} & \textbf{0.899} \\
Improve & 2.2\% & 4.2\% & 0.6\% & 1.2\% & 0.2\% \\
\hline
\end{tabular}
\label{table2}
}
\end{table}

\begin{table}[htbp]
\centering
\caption{Experimental results of the proposed approach \proposedmethod and baselines on the dataset of Tmall}
\resizebox{\linewidth}{!}{
\begin{tabular}{l|lllll}
\hline
Tmall & HR@5 & HR@10 & NDCG@5 & NDCG@10 & MRR@5 \\
\hline
FEARec & 0.143 & 0.157 & 0.118 & 0.144 & 0.128 \\
SASRecCPR & 0.145 & 0.163 & 0.105 & 0.119 & 0.103 \\
GRU4RecCPR & 0.149 & 0.154 & 0.106 & 0.111 & 0.108 \\
\hline
MAERec & 0.197 & 0.204 & 0.189 & 0.201 & 0.186 \\
SelfGNN & 0.171 & 0.190 & 0.164 & 0.185 & 0.153 \\
SURGE & 0.103 & 0.161 & 0.058 & 0.086 & 0.081 \\
\hline
PBAT & 0.216 & 0.278 & 0.209 & 0.263 & 0.213 \\
MBSRec & \underline{0.278} & 0.316 & \underline{0.214} & \underline{0.285} & \underline{0.217} \\
MBHT & 0.264 & \underline{0.349} & 0.202 & 0.229 & 0.181 \\
\hline
\textbf{PHKT} & \textbf{0.387} & \textbf{0.440} & \textbf{0.342} & \textbf{0.359} & \textbf{0.326} \\
Improve & 39.2\% & 26.1\% & 59.8\% & 26.0\% & 50.2\% \\
\hline
\end{tabular}
\label{table3}
}
\end{table}

\begin{table}[htbp]
\centering
\caption{Experimental results of the proposed approach \proposedmethod and baselines on the dataset of IJCAI}
\resizebox{\linewidth}{!}{
\begin{tabular}{l|lllll}
\hline
IJCAI & HR@5 & HR@10 & NDCG@5 & NDCG@10 & MRR@5 \\
\hline
FEARec & 0.116 & 0.138 & 0.113 & 0.125 & 0.111 \\
SASRecCPR & 0.124 & 0.140 & 0.112 & 0.126 & 0.106 \\
GRU4RecCPR & 0.120 & 0.151 & 0.086 & 0.102 & 0.094 \\
\hline
MAERec & 0.210 & 0.224 & 0.185 & 0.199 & 0.180 \\
SelfGNN & 0.201 & 0.217 & 0.183 & 0.205 & 0.178 \\
SURGE & 0.147 & 0.252 & 0.103 & 0.138 & 0.125 \\
\hline
PBAT & 0.259 & 0.291 & 0.247 & 0.286 & 0.245 \\
MBSRec & \underline{0.352} & \underline{0.389} & \underline{0.295} & \underline{0.316} & \underline{0.322} \\
MBHT & 0.244 & 0.329 & 0.178 & 0.206 & 0.157 \\
\hline
\textbf{PHKT} & \textbf{0.525} & \textbf{0.583} & \textbf{0.476} & \textbf{0.495} & \textbf{0.460} \\
Improve & 49.1\% & 49.9\% & 61.4\% & 56.7\% & 42.9\% \\
\hline
\end{tabular}
\label{table4}
}
\end{table}

\subsection{Parameter experiments}

This section investigates the impact of key hyperparameters in PHKT on model performance, including feature dimension, the number of KAN layers, and the initial behavior weight coefficient. Through systematic experiments on three datasets, we determine the optimal feature dimension, confirm its nonlinear relationship with performance, and further analyze its dependence on dataset characteristics.

\subsubsection{Effect of feature dimension}

\begin{figure*}[htbp]
  \centering
  \includegraphics[width=0.8\textwidth]{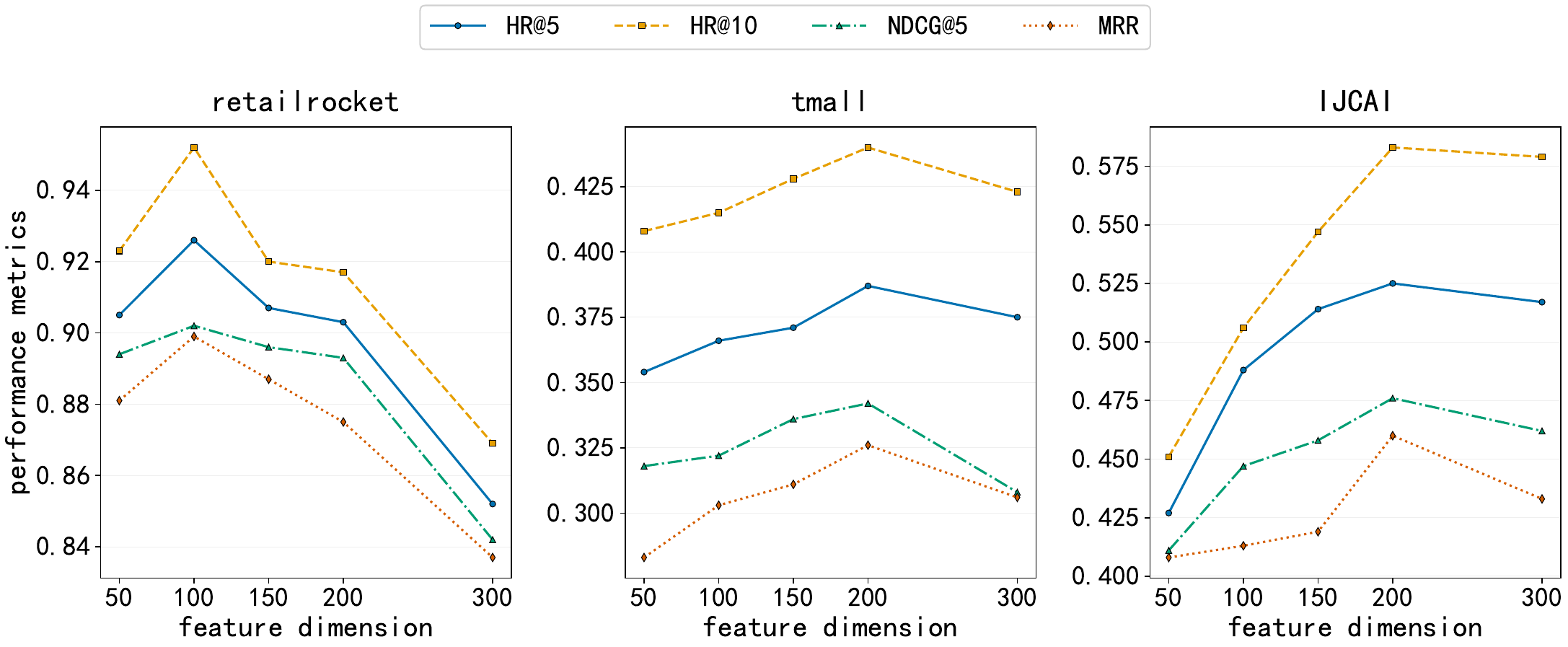}
  \caption{The impact of feature dimensionality on model performance}
  \label{fig:feature}
\end{figure*}

To analyze the effect of feature dimension on the performance of PHKT, we set the feature dimension to 50, 100, 150, 200, and 300, and conducted experiments on the RetailRocket, IJCAI, and Tmall datasets. As shown in Fig.~\ref{fig:feature}, as the feature dimension increases, the overall performance first improves and then declines. This indicates that overly low dimensions are insufficient to fully represent user and item information, whereas overly high dimensions may introduce redundant information and noise, thereby degrading model performance.

Specifically, on the RetailRocket dataset, the model achieves the best performance when the feature dimension is set to 100, where HR@10, HR@5, NDCG@5, and MRR reach 0.952, 0.926, 0.902, and 0.899, respectively. On the IJCAI and Tmall datasets, the optimal feature dimension is 200. When the feature dimension is further increased to 300, all evaluation metrics decline to varying degrees across the three datasets, indicating that increasing the feature dimension does not continuously improve performance.

Overall, the optimal feature dimension is highly dependent on the characteristics of the dataset. Compared with relatively simple datasets, those with more complex semantic relationships or higher sparsity generally require higher-dimensional representations to capture latent associations more effectively. These results suggest that the choice of feature dimension essentially reflects a trade-off between representation capacity and the risk of overfitting. Therefore, in practical applications, the feature dimension should be selected according to the properties of the dataset in order to achieve better model performance.

\subsubsection{Effect of the number of KAN layers}

\begin{figure*}[htbp]
  \centering
  \includegraphics[width=0.8\textwidth]{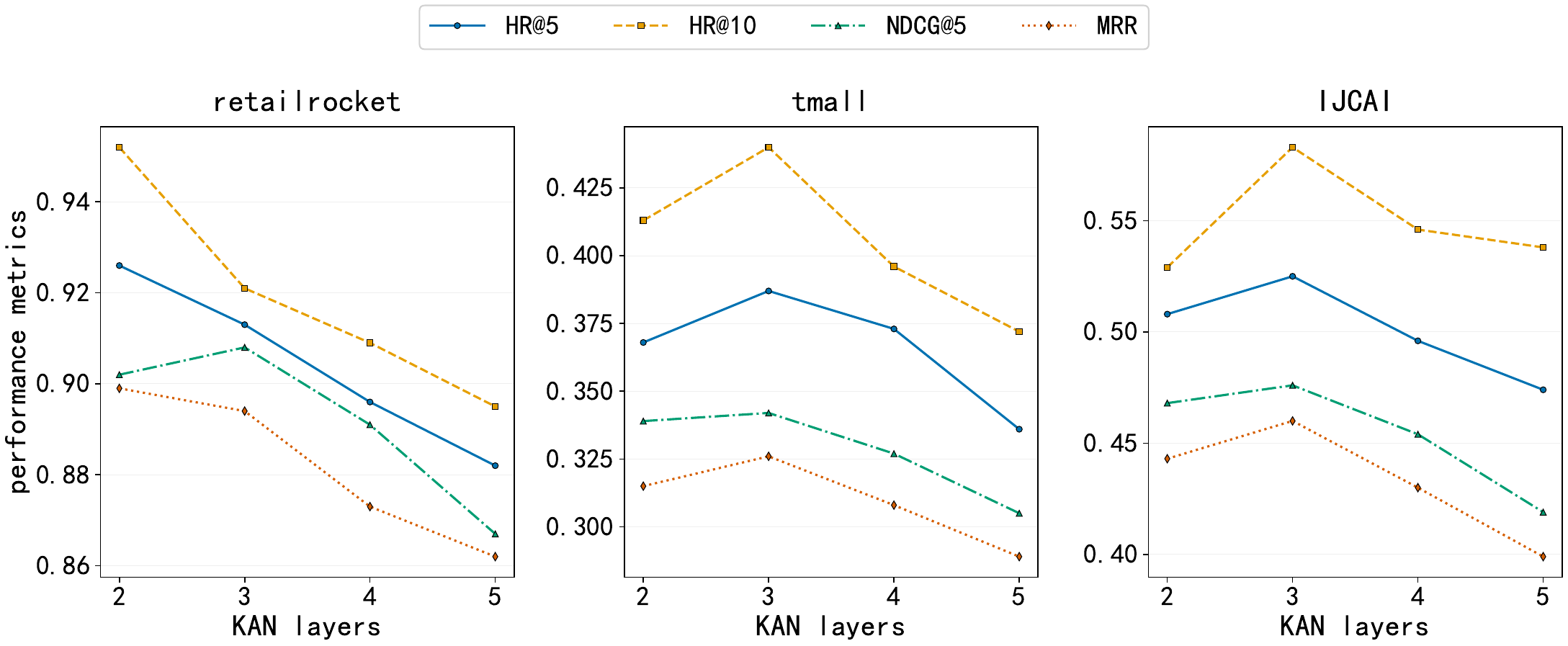}
  \caption{The impact of the number of KAN network layers on model performance}
  \label{fig:kannum}
\end{figure*}

To analyze the effect of the number of KAN layers on the performance of PHKT, we set the number of KAN layers to 2, 3, 4, and 5, and conducted experiments on the RetailRocket, IJCAI, and Tmall datasets. As shown in Fig.~\ref{fig:kannum}, as the number of KAN layers increases, the model performance generally exhibits either a declining trend or a pattern of first improving and then deteriorating. This indicates that a deeper network structure does not necessarily lead to continuous performance gains; instead, excessive depth may introduce overfitting and thus weaken the generalization ability of the model.

Specifically, on the RetailRocket dataset, the model achieves the best overall performance when the number of KAN layers is 2. On the IJCAI and Tmall datasets, the model performs best when the number of KAN layers is 3. When the number of KAN layers is further increased, all evaluation metrics decline to varying degrees across the three datasets, indicating that increasing network depth does not continuously improve performance. This result suggests that the optimal number of KAN layers differs across datasets.

Overall, the optimal number of KAN layers is also highly dependent on the characteristics of the dataset. Compared with relatively simple datasets, those with more complex semantic relationships or higher sparsity generally require a moderately deeper network to enhance feature extraction and representation ability. However, excessively deep network structures may still increase the risk of noise accumulation and overfitting. These results indicate that the choice of the number of KAN layers essentially reflects a trade-off between model expressiveness and generalization ability. Therefore, in practical applications, the network depth should be selected according to the characteristics of the dataset in order to achieve better recommendation performance.

\subsubsection{Effect of the initial behavior weight coefficient}

\begin{figure}
  \centering
  \includegraphics[
    width=0.6\textwidth,
    height=0.6\textheight,
    keepaspectratio
  ]{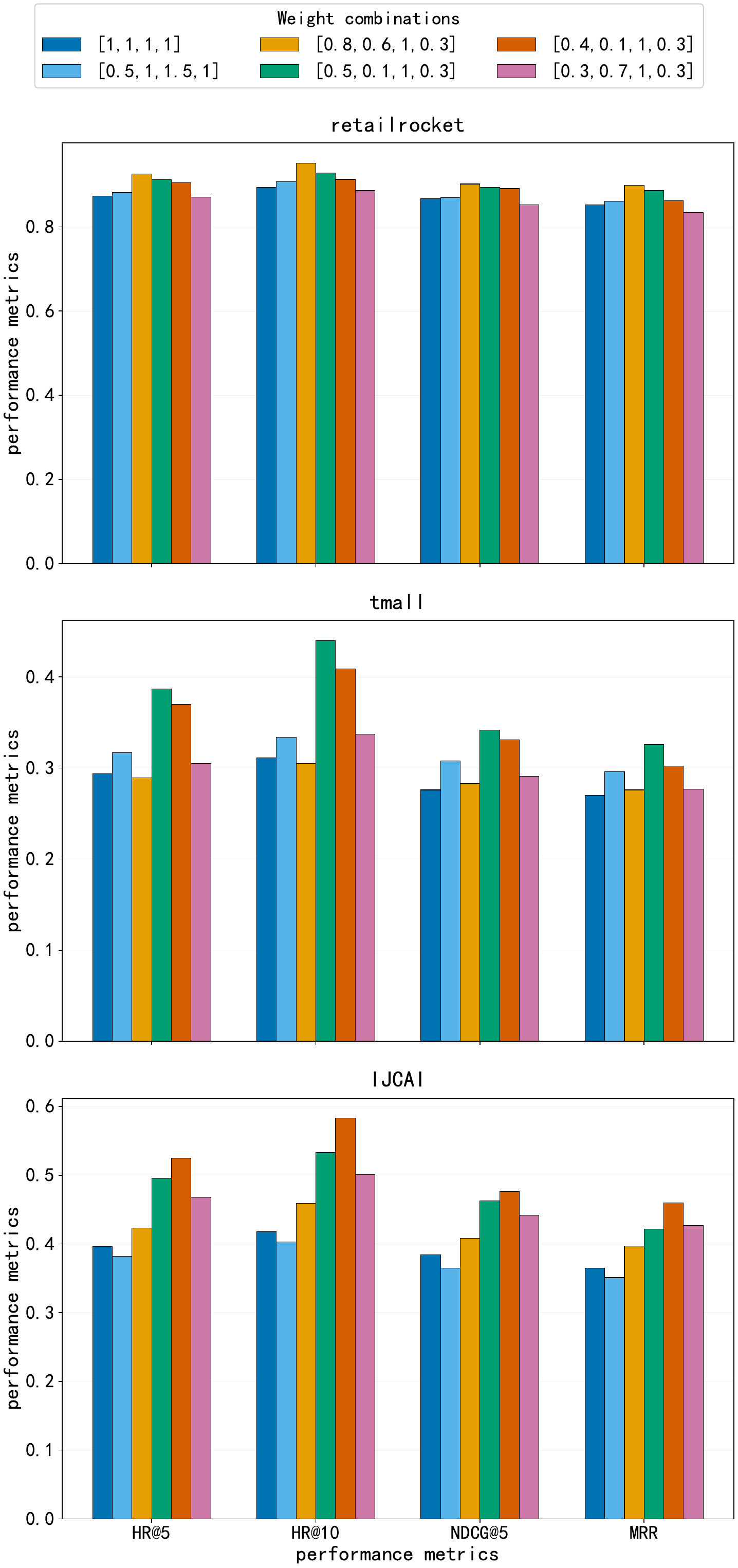}
  \caption{The impact of behavior weight combinations on model performance}
  \label{fig:weight}
\end{figure}

To analyze the effect of the initial behavior weight coefficient on the performance of PHKT, we conducted experiments with six different behavior weight combinations on the RetailRocket, IJCAI, and Tmall datasets. As shown in Fig.~\ref{fig:weight}, the results show that different weight configurations have a significant impact on model performance, and the optimal weight combination varies across datasets, indicating that the setting of behavior weights should be adjusted according to the characteristics of the data.

Specifically, on the RetailRocket dataset, the model achieves the best performance under the weight combination \([0.8,0.6,1,0.3]\); on the IJCAI dataset, the optimal weight combination is \([0.4,0.1,1,0.3]\); and on the Tmall dataset, the model performs best with \([0.5,0.1,1,0.3]\). Overall, the optimal configurations across all datasets adopt a differentiated weighting strategy, assigning the highest weight to purchase behavior, lower weight to browsing behavior, and moderate weights to cart and other behaviors. In contrast, the equal-weight setting \([1,1,1,1]\) yields relatively weaker performance, suggesting that a simple uniform weighting scheme cannot effectively capture the differences in information value among behavior types.

Overall, the optimal initial behavior weight coefficient is also highly dependent on the dataset. However, a consistent pattern can still be observed: purchase behavior, which most directly reflects users' true preferences, should be assigned the highest weight; browsing behavior, which contains more noise, should be assigned a lower weight; and cart and other behaviors should be assigned moderate weights. These results indicate that a reasonable behavior weighting strategy can effectively improve the performance of multi-behavior sequential recommendation. Therefore, in practical applications, the behavior weights should be carefully selected according to the characteristics of the dataset to achieve better recommendation performance.

\subsection{Ablation Studies}
To evaluate the contributions of the designed behavior aware Kolmogorov–Arnold Network Transformer (KANT) layer and the personalized hypergraph with dynamic weight in the proposed model \proposedmethod, we conducted ablation studies by systematically disabling these components.
Specifically, we designed two ablated variants of our \proposedmethod~model to isolate the effects of each module:
\begin{itemize}
\item \textbf{\proposedmethod-KANTrans.} In this variant, we replaced the KANT layer with the standard Transformer feedforward network, keeping all other settings unchanged. This allows us to assess the impact of the KANT component on sequential modeling performance.
\item \textbf{\proposedmethod-DynamicG.} In this variant, we removed the dynamic hypergraph module and used a static hypergraph instead, in order to verify the contribution and importance of personalized dynamic weighting in hypergraph construction.

The results of the ablation experiments on the RetailRocket and Tmall datasets are reported in Table~\ref{table5}.

\end{itemize}
\begin{table}[htbp]
\centering
\caption{Ablation Results}
\resizebox{\linewidth}{!}{
\begin{tabular}{c c c c c}
\hline
Dataset & Metrics & PHKT-w/o-KANT & PHKT-w/o-DHG & PHKT \\ \hline

\multirow{3}{*}{RetailRocket} 
& HR@5   & 0.908 & 0.914 & \textbf{0.926} \\
& NDCG@5 & 0.898 & \textbf{0.902} & \textbf{0.902} \\
& MRR@5    & 0.897 & 0.892 & \textbf{0.899} \\ \hline

\multirow{3}{*}{Tmall} 
& HR@5   & 0.338 & 0.365 & \textbf{0.387} \\
& NDCG@5 & 0.323 & 0.336 & \textbf{0.342} \\
& MRR@5    & 0.319 & 0.317 & \textbf{0.326} \\ \hline

\multirow{3}{*}{IJCAI} 
& HR@5   & 0.447 & 0.459 & \textbf{0.525} \\
& NDCG@5 & 0.435 & 0.438 & \textbf{0.476} \\
& MRR@5    & 0.411 & 0.422 & \textbf{0.460} \\ \hline
\end{tabular}
}
\label{table5}
\end{table}

We can observe that removing either the KANT layer or the dynamic hypergraph leads to a noticeable decline in all performance metrics across both datasets.

Specifically, on RetailRocket dataset, the full model (\proposedmethod) achieves HR@5 of 0.926, NDCG@5 of 0.902, and MRR@5 of 0.899. However, when the KANT layer is removed (\proposedmethod-w/o-KANT), these values decrease to 0.908, 0.898, and 0.897, respectively. Similarly, removing the dynamic hypergraph (\proposedmethod-w/o-DHG) results in HR@5 of 0.914, NDCG@5 of 0.902, and MRR@5 of 0.892.

Moreover, the results on Tmall dataset are consistent. The full model (\proposedmethod) yields HR@5 of 0.387, NDCG@5 of 0.342, and MRR@5 of 0.326. Disabling the KANT layer (\proposedmethod-w/o-KANT) reduces these to 0.338, 0.323, and 0.319, while removing the dynamic hypergraph (\proposedmethod-w/o-DHG) decreases the results to 0.365, 0.336, and 0.317, respectively.

The results on IJCAI dataset are consistent. The full model (\proposedmethod) yields HR@5 of 0.525, NDCG@5 of 0.476, and MRR@5 of 0.460. Disabling the KANT layer (\proposedmethod-w/o-KANT) reduces these to 0.447, 0.435, and 0.411, while removing the dynamic hypergraph (\proposedmethod-w/o-DHG) decreases the results to 0.459, 0.438, and 0.422.

Note that the performance drop caused by removing the KANT layer (\proposedmethod) is more pronounced than that caused by removing the dynamic hypergraph, especially on the IJCAI dataset. The results indicate that the KANT layer plays a particularly critical role in \proposedmethod by capturing long-range item dependencies and modeling complex sequential patterns in multi-behavior recommendation.
On the other hand, the dynamic hypergraph module also provides clear contributions, as evidenced by consistent improvements in all metrics when included. This results suggest that dynamically adjusting hypergraph weights according to user context and multi-behavior signals helps accurate modeling of user preference evolution.

Overall, these ablation results validate the contributions of both the KANT layer and the personalized dynamic hypergraph within \proposedmethod. These designed modules help \proposedmethod to fully capture both global sequence dependencies and personalized multi-behavioral relationships, boosting recommendation performance.





\section{Conclusions}
This paper proposes an innovative multi-behavior algorithm enhanced by personalized hypergraph with dynamic weight and behavior aware Kolmogorov–Arnold Network Transformer (KANT) layer, called \proposedmethod. The algorithm significantly improves the recommendation performance by deeply analyzing the multi-behavior dynamic weighted hypergraph and utilizing the differential information of user behaviors. Meanwhile, \proposedmethod uniquely combines a Kolmogorov–Arnold Network (KAN) layer with Transformer, which can effectively alleviate the lack of transparency of the feedforward layer and take advantage of the strong nonlinear fitting of KAN to cope with the challenges brought by diverse heterogeneous data. \proposedmethod excels in dynamically capturing user intentions and adjusting hypergraph weights, enabling it to more accurately analyze user behavior characteristics. The experimental results on three real world datasets verify the effectiveness of the proposed algorithm, which is superior to several advanced methods.

Future work could integrate advanced techniques, such as disentangled negative sampling~\cite{li2025disentangled}, and subgraph reasoning~\cite{shi2025llm} into the \proposedmethod framework to better leverage multi-behavior feedback signals and mitigate false negatives within its personalized hypergraph structure.

\section*{Acknowledgments}
This research is supported by the Natural Science Foundation of Zhejiang Province under Grant No.LZ25F020010 and ZCLMS25F0201, the CCF-NetEase ThunderFire Innovation Research Funding under Grant No. CCF-Netease202505, the National Natural Science Foundation of China under Grant No.62402151. The authors acknowledge the Supercomputing Center of Hangzhou Dianzi University for providing computing resources.

\balance

\printcredits

\bibliographystyle{cas-model2-names}

\bibliography{References}

@inproceedings{liao2024hypergraph,
  title={Hypergraph-Enhanced Contrastively Regularized Transformer for Multi-Behavior E-commerce Product Recommendation},
  author={Liao, Shuiying and Mok, PY},
  booktitle={2024 IEEE International Conference on Data Mining (ICDM)},
  pages={767--772},
  year={2024},
  organization={IEEE}
}

@article{vaswani2017attention,
  title={Attention is all you need},
  author={Vaswani, Ashish and Shazeer, Noam and Parmar, Niki and Uszkoreit, Jakob and Jones, Llion and Gomez, Aidan N and Kaiser, {\L}ukasz and Polosukhin, Illia},
  journal={Advances in neural information processing systems},
  volume={30},
  year={2017}
}

@inproceedings{rendle2010factorizing,
  title={Factorizing personalized markov chains for next-basket recommendation},
  author={Rendle, Steffen and Freudenthaler, Christoph and Schmidt-Thieme, Lars},
  booktitle={Proceedings of the 19th international conference on World wide web},
  pages={811--820},
  year={2010}
}

@article{hidasi2015session,
  title={Session-based recommendations with recurrent neural networks},
  author={Hidasi, Bal{\'a}zs and Karatzoglou, Alexandros and Baltrunas, Linas and Tikk, Domonkos},
  journal={arXiv preprint arXiv:1511.06939},
  year={2015}
}

@inproceedings{tang2018personalized,
  title={Personalized top-n sequential recommendation via convolutional sequence embedding},
  author={Tang, Jiaxi and Wang, Ke},
  booktitle={Proceedings of the eleventh ACM international conference on web search and data mining},
  pages={565--573},
  year={2018}
}

@inproceedings{li2019multi,
  title={Multi-interest network with dynamic routing for recommendation at Tmall},
  author={Li, Chao and Liu, Zhiyuan and Wu, Mengmeng and Xu, Yuchi and Zhao, Huan and Huang, Pipei and Kang, Guoliang and Chen, Qiwei and Li, Wei and Lee, Dik Lun},
  booktitle={Proceedings of the 28th ACM international conference on information and knowledge management},
  pages={2615--2623},
  year={2019}
}

@inproceedings{ou2025ls,
  title={LS-TGNN: Long and Short-Term Temporal Graph Neural Network for Session-Based Recommendation},
  author={Ou, Zhonghong and Zhang, Xiao and Zhu, Yifan and Lyu, Shuai and Liu, Jiahao and Ao, Tu},
  booktitle={Proceedings of the AAAI Conference on Artificial Intelligence},
  volume={39},
  number={12},
  pages={12426--12434},
  year={2025}
}

@inproceedings{lv2025dynamic,
  title={Dynamic Multi-Interest Graph Neural Network for Session-Based Recommendation},
  author={Lv, Mingyang and Liu, Xiangfeng and Xu, Yuanbo},
  booktitle={Proceedings of the AAAI Conference on Artificial Intelligence},
  volume={39},
  number={12},
  pages={12328--12336},
  year={2025}
}

@inproceedings{jin2020multi,
  title={Multi-behavior recommendation with graph convolutional networks},
  author={Jin, Bowen and Gao, Chen and He, Xiangnan and Jin, Depeng and Li, Yong},
  booktitle={Proceedings of the 43rd international ACM SIGIR conference on research and development in information retrieval},
  pages={659--668},
  year={2020}
}

@article{yan2025cd,
  title={cd-MBRec: Enhancing multi-behavior recommendation by explicitly modeling commonality and diversity},
  author={Yan, Cairong and Zhu, Ziyang and Zhang, Yiwei and Guan, Xiaopeng and Wan, Yongquan},
  journal={Intelligent Data Analysis},
  volume={29},
  number={2},
  pages={292--305},
  year={2025},
  publisher={SAGE Publications Sage UK: London, England}
}

@inproceedings{feng2019hypergraph,
  title={Hypergraph neural networks},
  author={Feng, Yifan and You, Haoxuan and Zhang, Zizhao and Ji, Rongrong and Gao, Yue},
  booktitle={Proceedings of the AAAI conference on artificial intelligence},
  volume={33},
  number={01},
  pages={3558--3565},
  year={2019}
}

@inproceedings{xia2021self,
  title={Self-supervised hypergraph convolutional networks for session-based recommendation},
  author={Xia, Xin and Yin, Hongzhi and Yu, Junliang and Wang, Qinyong and Cui, Lizhen and Zhang, Xiangliang},
  booktitle={Proceedings of the AAAI conference on artificial intelligence},
  volume={35},
  number={5},
  pages={4503--4511},
  year={2021}
}

@inproceedings{chang2024copy,
  title={To copy, or not to copy; that is a critical issue of the output softmax layer in neural sequential recommenders},
  author={Chang, Haw-Shiuan and Agarwal, Nikhil and McCallum, Andrew},
  booktitle={Proceedings of the 17th ACM International Conference on Web Search and Data Mining},
  pages={67--76},
  year={2024}
}

@article{xu2024fourierkan,
  title={Fourierkan-gcf: Fourier kolmogorov--arnold network--an effective and efficient feature transformation for graph collaborative filtering. arXiv: https://arxiv. org/abs/2406.01034 (2024)},
  author={Xu, Jinfeng and Chen, Z and Li, J and Yang, S and Wang, W and Hu, X and Ngai, ECH},
  journal={arXiv preprint arXiv:2406.01034},
  year={2024}
}

@article{park2024cf,
  title={Cf-kan: Kolmogorov-Arnold network-based collaborative filtering to mitigate catastrophic forgetting in recommender systems},
  author={Park, Jin-Duk and Kim, Kyung-Min and Shin, Won-Yong},
  journal={arXiv preprint arXiv:2409.05878},
  year={2024}
}

@inproceedings{liu2024multi,
  title={Multi-behavior generative recommendation},
  author={Liu, Zihan and Hou, Yupeng and McAuley, Julian},
  booktitle={Proceedings of the 33rd ACM International Conference on Information and Knowledge Management},
  pages={1575--1585},
  year={2024}
}

@article{choi2025hypergraph,
  title={Hypergraph temporal multi-behavior recommendation},
  author={Choi, Jooweon and Kwon, Junehyoung and Kim, Yeonghwa and Kim, Youngbin},
  journal={Engineering Applications of Artificial Intelligence},
  volume={145},
  pages={110112},
  year={2025},
  publisher={Elsevier}
}

@article{li2024hypergraph,
  title={Hypergraph-Enhanced Multi-interest Learning for multi-behavior sequential recommendation},
  author={Li, Qingfeng and Ma, Huifang and Jin, Wangyu and Ji, Yugang and Li, Zhixin},
  journal={Expert Systems with Applications},
  volume={255},
  pages={124497},
  year={2024},
  publisher={Elsevier}
}

@inproceedings{DBLP:conf/iclr/LiuWVRHS0T25,
  author       = {Ziming Liu and
                  Yixuan Wang and
                  Sachin Vaidya and
                  Fabian Ruehle and
                  James Halverson and
                  Marin Soljacic and
                  Thomas Y. Hou and
                  Max Tegmark},
  title        = {{KAN:} Kolmogorov-Arnold Networks},
  booktitle    = {The Thirteenth International Conference on Learning Representations,
                  {ICLR} 2025, Singapore, April 24-28, 2025},
  publisher    = {OpenReview.net},
  year         = {2025},
  url          = {https://openreview.net/forum?id=Ozo7qJ5vZi},
  bibsource    = {dblp computer science bibliography, https://dblp.org}
}

@article{somvanshi2025survey,
  title={A survey on kolmogorov-arnold network},
  author={Somvanshi, Shriyank and Javed, Syed Aaqib and Islam, Md Monzurul and Pandit, Diwas and Das, Subasish},
  journal={ACM Computing Surveys},
  volume={58},
  number={2},
  pages={1--35},
  year={2025},
  publisher={ACM New York, NY}
}

@article{liu2025weighted,
  title={Weighted nonlinear information extension based time series Kolmogorov--Arnold Network for industrial application with soft sensing},
  author={Liu, Guo-Yu and Zhu, Qun-Xiong and Zhang, Ning and He, Yan-Lin and Zhang, Ming-Qing and Xu, Yuan},
  journal={Engineering Applications of Artificial Intelligence},
  volume={160},
  pages={111719},
  year={2025},
  publisher={Elsevier}
}

@article{pan2025adaptive,
  title={Adaptive dissemination process in weighted hypergraphs},
  author={Pan, Qingtao and Wang, Zining and Wang, Haosen and Tang, Jun},
  journal={Expert Systems with Applications},
  volume={268},
  pages={126340},
  year={2025},
  publisher={Elsevier}
}

@article{liao2025fuzzy,
  title={Fuzzy clustering-based dual-channel contrastive learning for multi-behavior recommendation},
  author={Liao, Juan and Jantan, Aman and Liu, Zhe and Senapati, Tapan and Ulutagay, G{\"o}zde and Abualigah, Laith and Ahmed, Omed Hassan},
  journal={Engineering Applications of Artificial Intelligence},
  volume={157},
  pages={111381},
  year={2025},
  publisher={Elsevier}
}

@article{yu2026multi,
  title={Multi-behavioral recommendation algorithm based on decoupled graph convolution},
  author={Yu, Xu and Ding, Pengju and Yu, Jie and Lin, Junyu and Guo, Lei and Liu, Guanfeng and Xi, Liang},
  journal={Expert Systems with Applications},
  volume={298},
  pages={129618},
  year={2026},
  publisher={Elsevier}
}

@article{li2025disentangled,
  title={Disentangled progressive negative sampling for graph collaborative filtering recommendation},
  author={Li, Hewei and Zhang, Xin and Weng, He and Shen, Yingjie and Cai, Kangkai and Wang, Dongjing and Qin, Zhen and Deng, Shuiguang},
  journal={Knowledge-Based Systems},
  pages={114133},
  year={2025},
  publisher={Elsevier}
}

@article{shi2025llm,
  title={Llm-powered explanations: Unraveling recommendations through subgraph reasoning},
  author={Shi, Guangsi and Deng, Xiaofeng and Luo, Linhao and Xia, Lijuan and Bao, Lei and Ye, Bei and Du, Fei and Pan, Shirui and Li, Yuxiao},
  journal={Knowledge-Based Systems},
  pages={114307},
  year={2025},
  publisher={Elsevier}
}

@article{zhang2025multivariate,
  title={Multivariate Hawkes Spatio-Temporal Point Process with attention for point of interest recommendation},
  author={Zhang, Xin and Weng, He and Wei, Yuxin and Wang, Dongjing and Chen, Jia and Liang, Tingting and Yin, Yuyu},
  journal={Neurocomputing},
  volume={619},
  pages={129161},
  year={2025},
  publisher={Elsevier}
}

@article{li2024recent,
  title={Recent developments in recommender systems: A survey},
  author={Li, Yang and Liu, Kangbo and Satapathy, Ranjan and Wang, Suhang and Cambria, Erik},
  journal={IEEE Computational Intelligence Magazine},
  volume={19},
  number={2},
  pages={78--95},
  year={2024},
  publisher={IEEE}
}

@article{wang2025graph,
  title={Graph intention embedding neural network for tag-aware recommendation},
  author={Wang, Dongjing and Yao, Haojiang and Yu, Dongjin and Song, Shiyu and Weng, He and Xu, Guandong and Deng, Shuiguang},
  journal={Neural Networks},
  volume={184},
  pages={107062},
  year={2025},
  publisher={Elsevier}
}

@inproceedings{yang2024cascading,
  title={Cascading Multimodal Feature Enhanced Contrast Learning for Music Recommendation},
  author={Yang, Qimeng and Wang, Shijia and Guo, Da and Yu, Dongjin and Xiao, Qiang and Wang, Dongjing and Luo, Chuanjiang},
  booktitle={2024 IEEE International Conference on Data Mining (ICDM)},
  pages={905--910},
  year={2024},
  organization={IEEE}
}

@article{wang2023multi,
  title={Multi-view enhanced graph attention network for session-based music recommendation},
  author={Wang, Dongjing and Zhang, Xin and Yin, Yuyu and Yu, Dongjin and Xu, Guandong and Deng, Shuiguang},
  journal={ACM Transactions on Information Systems},
  volume={42},
  number={1},
  pages={1--30},
  year={2023},
  publisher={ACM New York, NY}
}

@article{yu2024mhaner,
  title={MHANER: A multi-source heterogeneous graph attention network for explainable recommendation in online games},
  author={Yu, Dongjin and Wang, Xingliang and Xiong, Yu and Shen, Xudong and Wu, Runze and Wang, Dongjing and Zou, Zhene and Xu, Guandong},
  journal={ACM Transactions on Intelligent Systems and Technology},
  volume={15},
  number={4},
  pages={1--23},
  year={2024},
  publisher={ACM New York, NY}
}

@article{chen2026gxnsrec,
  title={GXNSRec: Multi-behavior sequential recommender based on graph cross networks},
  author={Chen, Ruixin and Fan, Jianping and Wu, Meiqin and Cheng, Rui and Chai, Mingxuan},
  journal={Expert Systems with Applications},
  volume={297},
  pages={129387},
  year={2026},
  publisher={Elsevier},
  doi={10.1016/j.eswa.2025.129387}
}

@article{zhou2026hypergraph,
  title={Enhanced recommendation with hypergraph mixture of experts},
  author={Zhou, Zihao and Chen, Zhijun and Ma, Guofang and Lin, Zhenghong and Tan, Yanchao and Wang, Shiping and Yang, Carl},
  journal={Expert Systems with Applications},
  volume={297},
  pages={129333},
  year={2026},
  publisher={Elsevier},
  doi={10.1016/j.eswa.2025.129333}
}

@inproceedings{tan2016improved,
  author={Tan, Yong Kiam and Xu, Xinxing and Liu, Yong},
  title={Improved Recurrent Neural Networks for Session-based Recommendations},
  booktitle={Proceedings of the 1st Workshop on Deep Learning for Recommender Systems},
  year={2016},
  doi={10.1145/2988450.2988452}
}

@inproceedings{kang2018self,
  author={Kang, Wang-Cheng and McAuley, Julian},
  title={Self-Attentive Sequential Recommendation},
  booktitle={2018 IEEE International Conference on Data Mining (ICDM)},
  pages={197--206},
  year={2018},
  doi={10.1109/ICDM.2018.00035}
}

@inproceedings{du2023fearec,
  author={Du, Xinyu and Yuan, Huanhuan and Zhao, Pengpeng and Qu, Jianfeng and Zhuang, Fuzhen and Liu, Guanfeng and Liu, Yanchi and Sheng, Victor S.},
  title={Frequency Enhanced Hybrid Attention Network for Sequential Recommendation},
  booktitle={Proceedings of the 46th International ACM SIGIR Conference on Research and Development in Information Retrieval},
  pages={78--88},
  year={2023},
  doi={10.1145/3539618.3591689}
}

@inproceedings{chang2021surge,
  author={Chang, Jianxin and Gao, Chen and Zheng, Yu and Hui, Yiqun and Niu, Yanan and Song, Yang and Jin, Depeng and Li, Yong},
  title={Sequential Recommendation with Graph Neural Networks},
  booktitle={Proceedings of the 44th International ACM SIGIR Conference on Research and Development in Information Retrieval},
  pages={378--387},
  year={2021}
}

@inproceedings{ye2023maerec,
  author={Ye, Yuyu and Xia, Liang and Huang, Chao},
  title={Graph Masked Autoencoder for Sequential Recommendation},
  booktitle={Proceedings of the 46th International ACM SIGIR Conference on Research and Development in Information Retrieval},
  pages={321--330},
  year={2023}
}

@inproceedings{liu2024selfgnn,
  author={Liu, Yang and Xia, Liang and Huang, Chao},
  title={{SelfGNN}: Self-supervised Graph Neural Networks for Sequential Recommendation},
  booktitle={Proceedings of the 47th International ACM SIGIR Conference on Research and Development in Information Retrieval},
  pages={1609--1618},
  year={2024}
}

@inproceedings{yang2022multi,
  author={Yang, Yuchen and Huang, Chao and Xia, Liang and others},
  title={Multi-behavior Hypergraph-enhanced Transformer for Sequential Recommendation},
  booktitle={Proceedings of the 28th ACM SIGKDD Conference on Knowledge Discovery and Data Mining},
  pages={2263--2274},
  year={2022}
}

@inproceedings{su2023personalized,
  author={Su, Jiaqi and Chen, Chao and Lin, Zihan and others},
  title={Personalized Behavior-aware Transformer for Multi-behavior Sequential Recommendation},
  booktitle={Proceedings of the 31st ACM International Conference on Multimedia},
  pages={6321--6331},
  year={2023}
}

@inproceedings{elsayed2024multi,
  author={Elsayed, Sherein and Rashed, Ahmed and Schmidt-Thieme, Lars},
  title={Multi-behavioral Sequential Recommendation},
  booktitle={Proceedings of the 18th ACM Conference on Recommender Systems},
  pages={902--906},
  year={2024}
}

@inproceedings{sun2019bert4rec,
  title={BERT4Rec: Sequential Recommendation with Bidirectional Encoder Representations from Transformer},
  author={Sun, Fei and Liu, Jun and Wu, Jian and Pei, Changhua and Lin, Xiao and Ou, Wenwu and Jiang, Peng},
  booktitle={Proceedings of the 28th ACM International Conference on Information and Knowledge Management},
  pages={1441--1450},
  year={2019},
  publisher={ACM},
  doi={10.1145/3357384.3357895}
}


%
%

\end{document}